\documentclass[aps,pre,twocolumn,psfig,showpacs]{revtex4-1}
\usepackage{amsfonts}
\usepackage{amsmath}
\usepackage{graphicx}
\usepackage{bm}
\usepackage{amssymb}
\usepackage{times}
\usepackage{dcolumn}
\usepackage{cases}
\usepackage{txfonts}
\usepackage{mathtools}
\usepackage{setspace}

\begin{document}

\title{Quantum Monte Carlo Study on the Spin-1/2 Honeycomb Heisenberg Model with Mixing Antiferromagnetic and Ferromagnetic Interactions in External Magnetic Fields}
\author{Yi-Zhen Huang}
\affiliation{CAS Key Laboratory of Vacuum Physics, School of Physical Sciences, University of Chinese Academy of Sciences, P. O. Box 4588, Beijing
100049, China}
\author{Gang Su}
\email[Corresponding author. ]{Email: gsu@ucas.ac.cn}
\affiliation{Kavli Institute for Theoretical Sciences, and CAS Key Laboratory of Vacuum Physics, School of Physical Sciences, University of Chinese Academy of Sciences, Beijing 100049, China}

\begin{abstract}
 The continuous imaginary-time quantum Monte Carlo method with the worm update algorithm is applied to explore the ground state properties of the spin-1/2 Heisenberg model with antiferromagnetic (AF) coupling $J>0$ and ferromagnetic (F) coupling $J^{\prime}<0$ along  zigzag and armchair directions, respectively, on honeycomb lattice. It is found that by enhancing the F coupling $J^{\prime}$ between zigzag AF chains, the system is smoothly crossover from one-dimensional zigzag spin chains to a two-dimensional magnetic ordered state. In absence of an external field, the system is in a stripe order phase. In presence of uniform and staggered fields, the uniform and staggered out-of-plane magnetizations appear while the stripe order keeps in $xy$ plane, and a second-order quantum phase transition (QPT) at a critical staggered field is observed. The critical exponents of correlation length for QPTs induced by a staggered field for the cases with $J>0$, $J^{\prime}<0$ and $J<0$, $J^{\prime}>0$ are obtained to be $\nu=0.677(2)$ and $0.693(0)$, respectively, indicating that both cases belong to O(3) universality. The scaling behavior in a staggered field is analyzed, and the ground state phase diagrams in the plane of coupling ratio and staggered field are presented for two cases. The temperature dependence of susceptibility and specific heat of both systems in external magnetic fields is also discussed.
\end{abstract}

\maketitle

\section{Introduction}

Since the spin-1/2 antiferromagnetic (AF) Heisenberg model is believed to be capable of describing the undoped precursors of high temperature superconducting cuprates, it has attracted intensive attention in condensed matter and statistical physics. Through extensive explorations both theoretically and experimentally in the past decades, many properties of this model have been exposed, and  a great deal of advances have been achieved. However, as the complexity occurs intrinsically in many-body systems, there still remain a lot of ambiguities remaining to be investigated. For instance, by searching for exotic states of matter or studying quantum phase transitions (QPTs), people usually invoke this model with different interactions on various lattices as prototypes. To name but a few, quantum spin liquid is thought to exist in spin-1/2 AF Heisenberg models on lattices with geometrical frustrations, but its nature is still under active debate \citep{white, Depenbrock, T.H.Han, Y.Iqbal, Xie, T.Liu1, T.Liu2, T.Liu3};
whether exotic phase transitions beyond the traditional Landau-Ginzburg-Wilson framework \cite{awsandvik} exist or not were also discussed by introducing more complex interactions or by tuning the spatial anisotropy in coupling strength; and so on.

It has been shown that the spin-1/2 AF Heisenberg system on square lattice displays $N\acute{e}el$ order in ground state with the staggered magnetic moment per site $m_{s} \cong 0.3075$ by quantum Monte Carlo (QMC) \cite{runge} and $m_{s} \cong 0.3034$ by the spin-wave theory \cite{oguchi}, while the one-dimensional (1D) AF spin chain is magnetically disordered with gapless excitations.  The spin-1/2 AF Heisenberg chain is critical according to the exact result of Bethe ansatz or the results of, e.g., spin-wave theory with random phase approximation \cite{rosner} and mean-field approaches \cite{schulz}, which is also confirmed by using the multi-chain mean-field method associated with SSE Monte Carlo algorithm \cite{Sandvik}. The experiments on quasi-1D spin-1/2 AF chains such as $Sr_{2}CuO_{3}$ and $Ca_{2}CuO_{3}$ \cite{kojima} trigger an interesting question: how does the AF long-range order in two-dimensional (2D) lattice (like square lattice) in ground state develop from coupled 1D spin chains with increasing inter-chain interactions? There are a number of studies to tackle this issue and determine the critical inter-chain coupling ratio $R_{c}=J_{\perp}/J$: the spin wave theory gives $R_{c}=0.034$ \cite{sakai}, one-loop renormalization group analysis on an effectively spatially anisotropic nonlinear sigma model yields $R_{c}=0.047$ \cite{castro}, the series expansion numerical techniques bound $R_{c}$ upper to 0.02 \cite{affleck}; whereas some self-consistent calculations \cite{aoki} and exact diagnalization \cite{ihle} predict it as high as 0.15 and $0.1\sim0.2$, respectively. Therefore, this question still calls on further more accurate explorations.

Alternatively, one can also study 2D anisotropic Heisenberg AF models with different bond interactions to observe crossover behaviors by tuning the bond interactions. For instance, such attempts were made on honeycomb lattice with the dimer pairs pinned on the armchair bonds by using the methods of tensor renormalization group  \cite{wei} and QMC \cite{Ugerber},  where it is found that there is a QPT of classical O(3) universality from a disordered dimer phase to quantum $N\acute{e}el$ order at a critical inter-dimer AF interaction. In our previous work \cite{yizhen}, we replaced the inter-dimer AF couplings by ferromagnetic (F) interactions along zigzag directions on honeycomb lattice, and found that there is also a phase transition from a dimerized phase to a stripe phase. The scaling behaviors were analyzed, and the coupling parameters of two compounds were estimated by comparing our QMC calculated results.

In contrast to this previous work \cite{yizhen}, where the interactions are supposed to be F along zigzag direction and AF along armchair direction, for the completeness of the study, in this paper we shall consider the spin-1/2 Heisenberg model with mixing AF interaction ($J$) along zigzag direction and F interaction ($J'$) along armchair direction on honeycomb lattice (Fig. 1) in magnetic fields. It should be remarked here that the present system (we refer to Case A later) is quite different from that considered in Ref. \onlinecite{yizhen} (we refer to Case B later), and the two systems cannot be transformed mutually by simply using a unitary transformation. This observation is confirmed by our QMC studies, where we observe that a small inter-chain F interaction could make the 1D disordered state smoothly crossover to a 2D spin ordered state. In presence of uniform and staggered fields, the uniform and staggered magnetizations in $z$ direction appear while a stripe order keeps in $xy$ plane, and a second-order QPT at a critical staggered field is observed.
Unlike there is a zero magnetization plateau in the honeycomb spin ladder with AF legs and F rungs in $m_{z}\sim\emph{h}$ curves \cite{yang}, no zero magnetization plateau exists for both cases with $J'=-J$ owing to the appearance of spin order. The phase diagram in a staggered field is also presented.

This paper is organized as follows. In Sec. \uppercase\expandafter{\romannumeral2}, we shall give the model Hamiltonian, calculational method and definitions of several physical quantities; the crossover behavior from 1D to 2D in  absence of a magnetic field is discussed in Sec. \uppercase\expandafter{\romannumeral3}; the magnetization in presence of uniform and staggered external magnetic fields is presented in Sec. \uppercase\expandafter{\romannumeral4}; the finite-size scaling analysis is given in Sec. \uppercase\expandafter{\romannumeral5}; Sec. \uppercase\expandafter{\romannumeral6} shows the phase diagrams of two systems for a comparison; the temperature dependence of specific heat and susceptibility in magnetic fields is discussed in  Sec. \uppercase\expandafter{\romannumeral7}; and finally, a summary is given.

\begin{figure}
\includegraphics[width=0.42\textwidth]{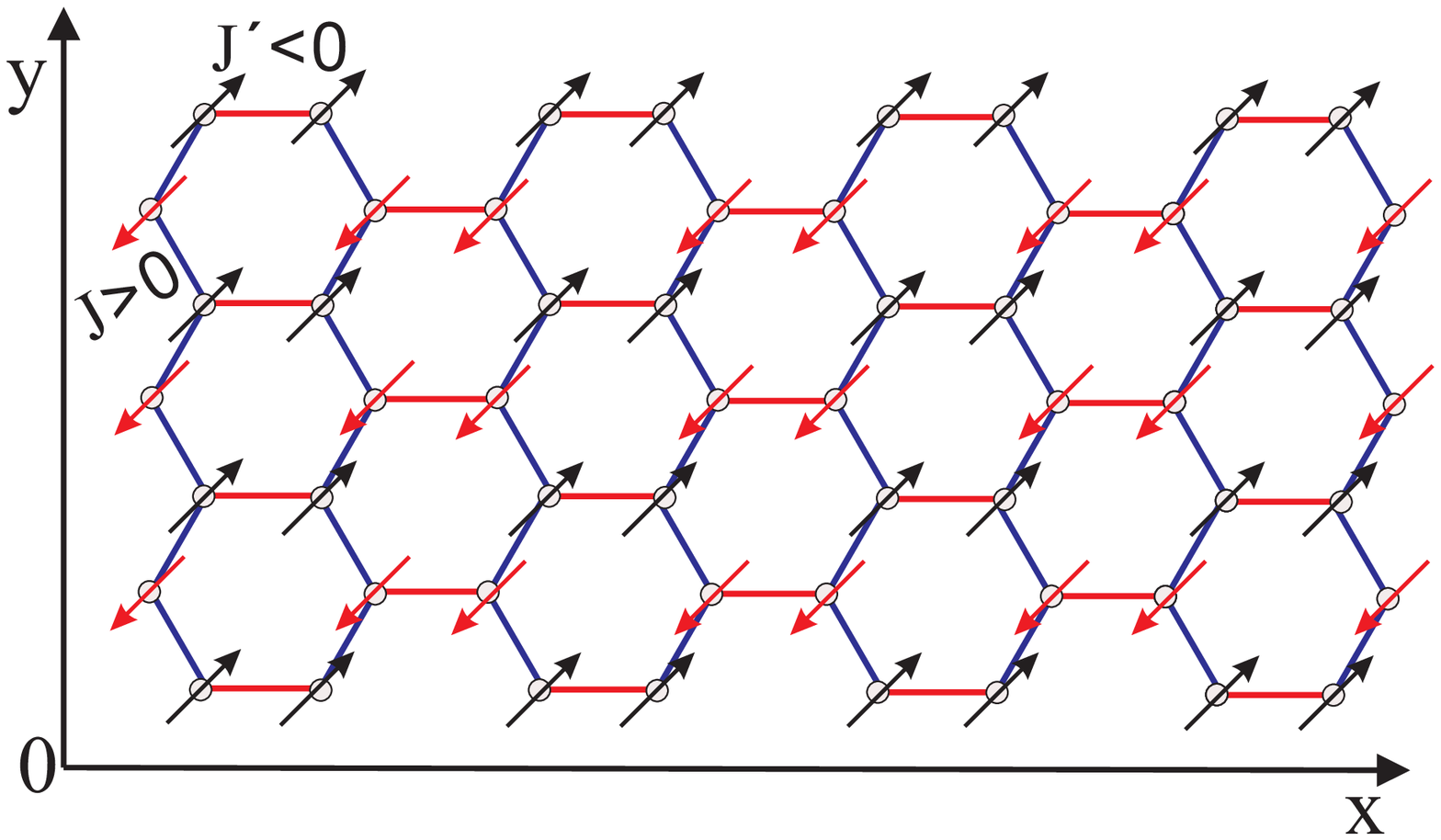}
\caption{(Color online) The spin-1/2 Heisenberg model on honeycomb lattice with antiferromagnetic (AF) and ferromagnetic (F) interactions between nearest neighbor spins (indicated by arrows) along zigzag (blue bonds along $y$ direction) and armchair (red bonds along $x$ direction) chains, respectively.}
\label{interaction-pattern}
\end{figure}

\section{Model, Method and Definitions}

\subsection{Model}

By using the continuous imaginary time QMC with worm update algorithm, we shall study the spin-1/2 Heisenberg model on honeycomb lattice with mixing AF and F interactions along zigzag and armchair directions, respectively, as depicted in Fig. \ref{interaction-pattern}, in presence of uniform or staggered magnetic fields. The Hamiltonian of the system is
\begin{eqnarray}
\begin{aligned}
H=J \!\sum_{\mathclap{\langle ij\rangle_{ZZ}}} \!\bold{S}_{i}\! \cdot\! \bold{S}_{j}+J^{\prime}\!\sum_{\mathclap{\langle ij\rangle_{AM}}}\!\bold{S}_{i}\!\cdot\! \bold{S}_{j}-\emph{h}\!\sum_{l=1}^{N}\!{\bold{S}_{l}^{z}}-\emph{h}_{s}\! \sum_{l=1}^{N}\!{(-1)^{l}\bold{S}_{l}^{z}},
\end{aligned}
\label{eq-hamiltonian}
\end{eqnarray}
where $\bold{S}_{i}$ is the spin-1/2 operator at $i$-th site,  $\langle ij\rangle_{ZZ}$ and $\langle ij\rangle_{AM}$ denote nearest neighbours along zigzag and armchair directions, respectively,  $J>0$ and $J^{\prime}<0$ are corresponding coupling constants, $h$ and  $h_s$ are the external uniform and staggered magnetic fields, respectively. We define the coupling ratio $\alpha_{1}=\frac{J^{\prime}}{J}$ for a later use.  For convenience, we also mark the armchair and zigzag directions by $x$ and $y$ directions, respectively. The lattice size is $N=L_{x}\times L_{y}$ with $L_{x(y)}$ the length of $x$ $(y)$ direction.

From Sec. \uppercase\expandafter{\romannumeral4} on, we shall also make comparisons between the present system and the system with $J<0$ and $J^{\prime}>0$ in the previous work \cite{yizhen} in a staggered field, where the coupling ratio is defined as $\alpha_{2}=\frac{J}{J^{\prime}}$.

\subsection{Method}

We shall use the continuous time QMC with worm update algorithm to study the system under consideration. This algorithm expands the partition function of system as a summation of path integrals with continuous loops under Fork states representation with $\{|S_{i}^{z}\rangle\}$ as the basis in interaction picture by
\begin{eqnarray}
\begin{aligned}
Z&=Tr(e^{-\beta H})\\
&=Tr(e^{-\int_{0}^{\beta}d\tau H})\\
&=\sum_{n=0}^{\infty}(-1)^{n}\!Tr\{e^{-\beta H_{0}}\int_{0}^{\beta}\!d\tau_{n}...\int_{0}^{\tau_{2}}\!d\tau_{1}(H_{In}H_{I(n-1)}..H_{I1})\},
\end{aligned}
\label{partition-function}
\end{eqnarray}
where $\beta=\frac{1}{k_{B}T}$, the inverse temperature, acting as the length of the imaginary time in the simulation, $k_{B}=1$ the Boltzman constant, $H_{0}$ stands for the interaction between spins along the $z$ direction
\begin{eqnarray}
\begin{aligned}
H_{0}\!=\!J\! \sum_{\mathclap{\langle ij\rangle_{ZZ}}}S^{z}_{i}S^{z}_{j}+J^{\prime}\!\sum_{\mathclap{\langle ij\rangle_{AM}}}S^{z}_{i}S^{z}_{j}-\emph{h}\!\sum_{l=1}^{N}{S_{l}^{z}}-\emph{h}_{s}\! \sum_{l=1}^{N}{(-1)^{l}S_{l}^{z}},
\end{aligned}
\label{diagonal}
\end{eqnarray}
and $H_{I}$ is the hopping term in the $xy$ plane of spin space
\begin{eqnarray}
\begin{aligned}
H_{I}=\frac{J}{2}\sum_{\langle ij\rangle_{ZZ}}(S^{+}_{i}S^{-}_{j}+S^{-}_{i}S^{+}_{j})+\frac{J^{\prime}}{2}\sum_{\langle ij\rangle_{AM}}(S^{+}_{i}S^{-}_{j}+S^{-}_{i}S^{+}_{j}).
\end{aligned}
\label{off-diagonal}
\end{eqnarray}

In this framework, the object sampled during executing the algorithm is each term in Eq. (\ref{partition-function}), and the integration is concretized to several configurations with special localizations of off-diagonal terms in the imaginary time axis. In lattice space and imaginary time coordination, such configurations are graphed as multiple worldlines with only continuous loops. By introducing the kink pair $S^{+}_{i\tau_{1}}S^{-}_{i\tau_{2}}$, called a worm, a partition function configuration is switched into a Green function configuration. The hopping of the worm ends ($S^{+}_{i\tau}$ or $S^{-}_{i\tau}$) along the imaginary time or real space direction realizes the sampling of the Green functions and the ends' annihilation finishes an update from an old Z configuration to a new one \cite{prokofef}. The big difference between the partition function configuration and the Green function configuration is that the latter has an extra discontinuous worldline, i.e., the worm. This method extends the sampling space and could be used to calculate the winding number directly.

\subsection{Definitions}

Before we proceed further, we first give the definitions of relevant physical quantities that will be used later. As the calculations based on the QMC method are usually associated with the finite-size systems, where the spin O(3) rotational symmetry remains in a finite system, we should determine the order parameters by calculating the corresponding square values combined with a size extrapolation. The staggered magnetization per site is defined via the following expression
\begin{eqnarray}
\begin{aligned}
\langle m^{2}_{s}\rangle&=\langle\{\frac{1}{N}\sum_{j}^{N}(-1)^{j}(\mathbf{S}_{j}^{x}
+\mathbf{S}_{j}^{y}+\mathbf{S}_{j}^{z})\}^{2}\rangle\\
&=3\langle(\frac{1}{N}\sum_{j}^{N}(-1)^{j}\mathbf{S}_{j}^{z})^{2}\rangle\\
&=\frac{3}{2}\frac{1}{N}\sum_{0}^{N}f(r)<S_{0}^{+}S_{r}^{-}>,\\
\end{aligned}
\label{eq-stripe-magnetization}
\end{eqnarray}
where $f(r)=1$ if $S_{0}^{+}$ and $S_{r}^{-}$ are both in the even or odd zigzag (armchair) bonds, otherwise $f(r)=-1$.

The staggered magnetization per site in the $xy$ plane in presence of unform or staggered field can be studied through
\begin{eqnarray}
\begin{aligned}
\langle m_{\perp}^{2}\rangle&=\frac{1}{N^{2}}\sum_{n_{1}=0,n_{2}=0}^{n_{1}=N,n_{2}=N}f(r)\langle(S_{n_{1}}^{x}S_{n_{2}}^{x}+S_{n_{1}}^{y}S_{n_{2}}^{y})\rangle\\
&=\frac{1}{N}\sum_{0}^{N}f(r)\langle S_{0}^{+}S_{r}^{-}\rangle.\\
\end{aligned}
\label{eq-horizontal-magnetization}
\end{eqnarray}

The uniform magnetization per site is defined by
\begin{eqnarray}
\begin{aligned}
m_{z}=\langle\frac{1}{N\beta}\sum_{i=1}^{N}\int_{0}^{\beta}S_{i\tau}^{z}d\tau\rangle.
\end{aligned}
\label{eq-net-magnetization}
\end{eqnarray}

The uniform magnetic susceptibility is given by
\begin{eqnarray}
\chi_{u}\!&=&\!\frac{1}{N\beta}\{\!\sum_{ij}\!\langle\int_{0}^{\beta}\!d\tau_{1}d\tau_{2} S_{i\tau_{1}}^{z}S_{j\tau_{2}}^{z}\rangle \nonumber \\
&-&\langle\!\int_{0}^{\beta}\!d\tau_{1}S_{i\tau_{1}}^{z}\rangle\langle\!\int_{0}^{\beta}\!d\tau_{2}S_{j\tau_{2}}^{z}\rangle\}.
\label{eq-susceptibility}
\end{eqnarray}

The staggered magnetization per site in $z$ direction will be calculated by
\begin{eqnarray}
\begin{aligned} m_{z}^{s}=\langle\frac{1}{N\beta}\sum_{i=1,j=1}^{i=L_{x},j=L_{y}}(-1)^{i+j}\int_{0}^{\beta}S_{ij(\tau)}^{z}d\tau\rangle.
\end{aligned}
\label{eq-staggered-magnetization}
\end{eqnarray}

The spin stiffness $\rho$ is obtained by the fluctuation of winding numbers \cite{sandvik}
\begin{eqnarray}
\begin{aligned}
\rho_{\theta} &=\frac{\partial^{2}\Omega}{\partial^{2}\Phi}=\frac{1}{\beta}\langle W_{\theta}^{2}\rangle\\
&=\frac{1}{\beta}\langle [(N_{\theta}^{+}-N_{\theta}^{-})/L]^{2}\rangle,
\label{eq-spinstiffness}
\end{aligned}
\end{eqnarray}
where $\Omega$ is the free energy, $\Phi$, $W_{\theta}$, $N_{\theta}^{+}$ and $N_{\theta}^{-}$ are twisted angle at the boundaries, the winding number, number of sites for spin $\uparrow$ hopping along the positive and negative $\theta$ directions, respectively. It is noted that the spin stiffness $\rho$ in Eq. (\ref{eq-spinstiffness}) has its counterpart in a boson system \cite{jasnow,josephson}, the superfluid density, which characterizes an off-diagonal long range order.

\section{Crossover from 1D to 2D}

Now let us consider the crossover behavior from 1D to 2D by altering the coupling ratios of the present system. During the calculations, the coupling ratio is changed to the value as low as $\alpha_{1}=-0.010$.

\begin{figure}
\includegraphics[width=0.45\textwidth]{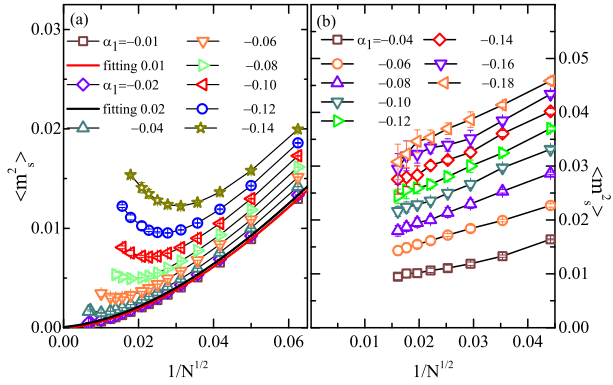}
\caption{(Color online) The size extrapolation of the order parameter $\langle m^{2}_{s}\rangle$ versus $\frac{1}{\sqrt{N}}$ for (a) $L_{x}=L_{y}=L$ and (b) $L_{y}=8L_{x}$. $N=L_{x}*L_{y}$ is the total number of lattice sites, and the inverse temperature is set to be $\beta=2*\sqrt{N}/J$ for (a) and $\beta=\sqrt{N}/J$ for (b). In (a), $\langle m^{2}_{s}\rangle$ exhibits a minimum for every $\alpha_{1}$. When $L_{y}\gg L_{x}$, $\langle m^{2}_{s}\rangle$ decreases almost linearly against $1/\sqrt{N}$ in (b). The errors that are visible are of order $10^{-3}$ and the invisible ones are of $10^{-4}$ at least. }
\label{order-size}
\end{figure}

 We first take $L_{x}$ and $L_{y}$ to be equal and the inverse temperature to be size-dependent $\beta=2*\sqrt{N}/J$. Fig. \ref{order-size} (a) shows the size extrapolation of $m^{2}_{s}$ versus $1/\sqrt{N}$ for various coupling ratio $\alpha$. It is seen that as the system size increases, $\langle m^{2}_{s}\rangle$ first decreases and then increases, leaving a minimum at a finite size for each $\alpha$, and in this case, it shows a non-monotonic behavior, and therefore a size extrapolation is impossible. It is known that the finite size gap in the zigzag chain is $\bigtriangleup (L_{y})\sim\frac{1}{L_{y}}$. When energy scales in the two directions are compatible, $\bigtriangleup (L_{y})\sim\rho_{x}$, where $\rho_{x}$ scales the energy in the armchair direction, the system
 crossovers from 1D to 2D \cite{Sandvik}, and in this way, the size extrapolation for the order parameter square is meaningful. In order to make a reasonable size extrapolation, we should take $L_{y}\gg L_{x}$ for small $|\alpha_{1}|$.

We carry out simulations on lattices with $L_{y}=8L_{x}$ for $\alpha_{1}$ from -0.04 to -0.18, where the inverse temperature is taken as $\beta=\sqrt{N}/J$. The size extrapolations to the square root of the total number of lattice sites $N=L_{x}*L_{y}$ are shown in Fig. \ref{order-size} (b). It can be observed that $\langle m^{2}_{s}\rangle$ decreases almost linearly against $1/\sqrt{N}$. By doing a polynomial fitting of order two to the data for $\alpha=-0.04$, the size extrapolation of $\langle m^{2}_{s}\rangle$ gives 0.0100(3), suggesting that the system is magnetically ordered.

In Fig. \ref{order-size} (a), the fitting curves for $\alpha_{1}=-0.02$ is very close to that for $\alpha_{1}=-0.01$ till the thermodynamic limit is reached, where the two curves do not have crossings, implying that the system is in a spin ordered state when $|\alpha_{1}|$ is larger than 0.01. The result $\lim_{L\rightarrow\infty}\langle m^{2}_{s}\rangle\rightarrow0.0100(3)$ for $\alpha=-0.04$, confirms the above observation. Therefore,  upon tuning a very small inter-chain F interaction between the AF zigzag spin chains, the system immediately undergoes a crossover from a disordered 1D state to a 2D spin ordered state.

\section{Magnetization in magnetic fields}

We now consider the effects of external uniform and staggered fields on magnetization and susceptibility of the system under interest.

\subsection{Presence of a uniform field}

In the presence of a uniform field $\emph{h}$,  the uniform magnetization $m_{z}$, the staggered magnetization in $xy$ plane $m^{s}_{\bot}=\sqrt{\langle m^{2}_{\bot}\rangle}$, and the uniform magnetic susceptibility $\chi_{u}$, have been calculated for $\alpha_{1}=-0.3$ on lattices with $L_{x}=L_{y}=32,36$, and $\beta=100/J$ for $L_{x}=32$ and $108/J$ for $L_{x}=36$.

\begin{figure}
\includegraphics[width=0.45\textwidth]{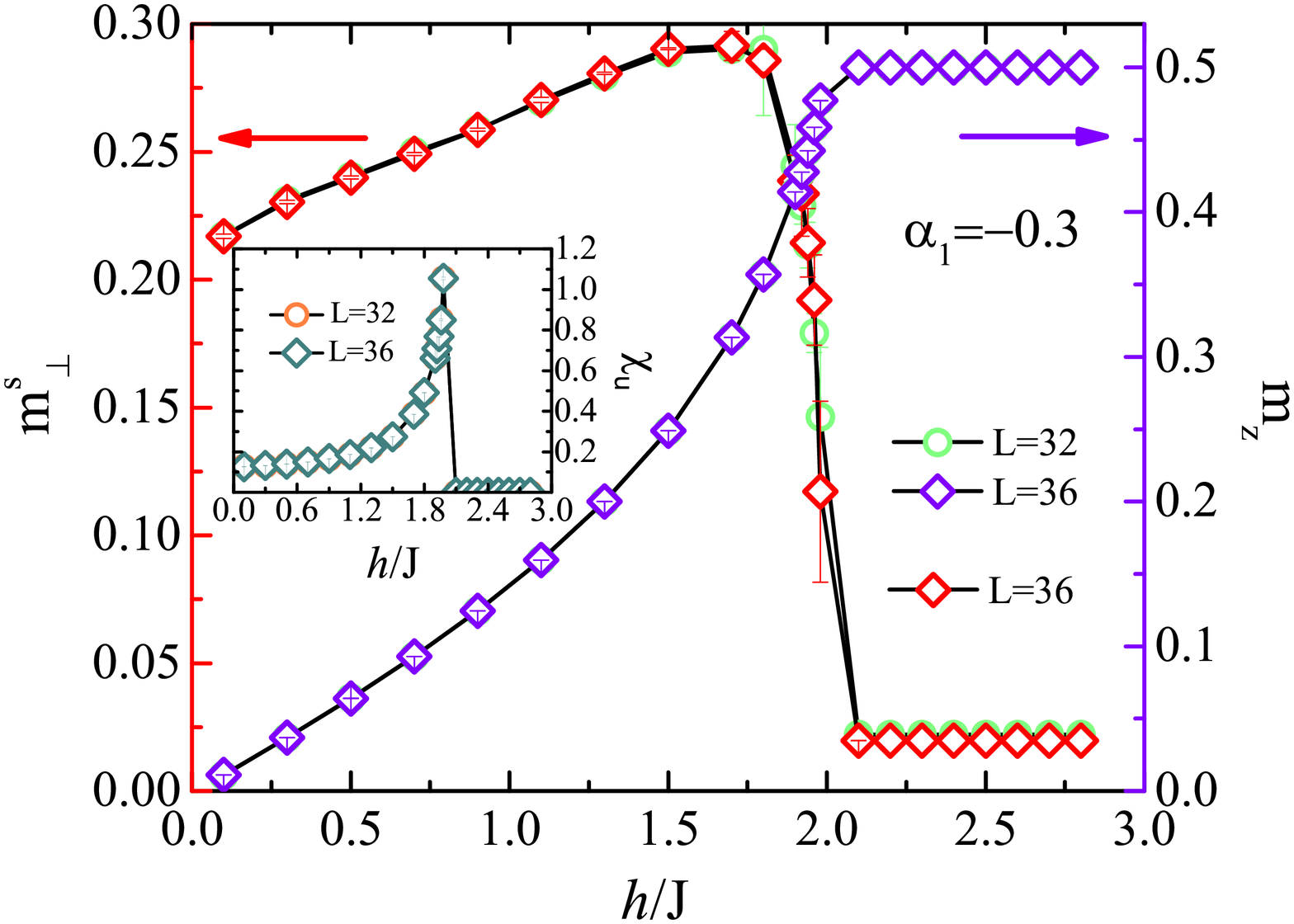}
\caption{(Color online) $m^{s}_{\perp}$ $m_{z}$, and $\chi_u$ vs uniform magnetic field $\emph{h/J}$ for different lattice sizes $L_{x}=L_{y}=32$ with $\beta=100/J$ and 36 with $\beta=108/J$ for $\alpha_{1}=-0.3$. Inset is the susceptibility $\chi_u$ as a function of $\emph{h/J}$. Around $\emph{h/J} = 2$ there is a second-order phase transition between a canted stripe phase and a polarized phase. Except the points whose error bar is visible, the rest data are calculated with accuracy of at least $10^{-3}$.}
\label{uniform-filed}
\end{figure}

Fig. \ref{uniform-filed} shows the field dependence of $m_{z}$, $m^{s}_{\bot}$ and $\chi_{u}$ for $\alpha_{1}=-0.3$. One may see that, in a weak field, $m_{z}$ increases from zero slowly and goes to saturation when $\emph{h/J}$ approaches to 2.0, while $m^{s}_{\bot}$ enhances from a finite value to a peak and then declines sharply around $\emph{h/J}=2.0$, and reaches zero when $\emph{h/J}>2.0$. The susceptibility $\chi_{u}$ in the inset displays a sharp peak at $\emph{h/J}=2.0$, indicating a second-order in-plane phase transition, and being consistent with the observation in magnetic curves. When $\emph{h}$ is applied, the $z$-component (out of plane component) of magnetic moments begins to develop with the decay of the in-plane $m^{s}_{\bot}$; when $\emph{h/J} \geq 2$, the system is fully polarized and the transverse component is totally suppressed. We call such a phase before fully polarized the canted stripe phase, in which  $m^{s}_{\bot}>0$, and $m_{z}>0$.

In contrast, for the system considered in Ref.  \onlinecite{yizhen} where $J<0$ and $J^{\prime}>0$, there is a phase transition at $\alpha_{2}=\frac{J}{J^{\prime}}\simeq-0.93$ from a disorder dimer phase to an ordered stripe phase, where the spin alignments are parallel along the same zigzag line and antiparallel along the armchair direction. $m_{z}$, $\tilde{m}^{s}_{\bot}$ (the staggered magnetization in $xy$ plane for the system explored in Ref.  \onlinecite{yizhen}) and $\chi$ for the ordered phase look like the ones of this present system. When it is in the dimer phase for $\alpha_{2}=-0.6$, the zero magnetization plateau appears in $m_{z}$, $\tilde{m}^{s}_{\bot}$ and $\chi$. Since the present system has an in-plane long range order (${m}^{s}_{\bot} \neq 0$) in the ground state, the excitation is gapless with Goldstone bosons, and there is no zero magnetization plateau in the magnetic curve (Fig. 3).

\subsection{Presence of a staggered field}

In this subsection, we shall investigate the magnetic curves the present system (we call Case A) and the system discussed in Ref. \onlinecite{yizhen} (we call Case B) in a staggered field $\emph{h}_{s}$ (while keeping $\emph{h}=0$). We study the staggered magnetization $m^{s}_{z}$ (Eq. (\ref{eq-staggered-magnetization})) and the magnetization square in the $xy$ plane $\langle m^{2}_{\bot}\rangle$ under a field $\emph{h}_{s}$ for $\alpha_{1}=-1.0$, and $\langle\tilde{m}^{2}_{\bot}\rangle$ for $\alpha_{2}=-1.0$, respectively. The simulations are performed on lattices with $L_{x}=L_{y}=$30, 36, 42, 48, and $\beta=100/J(J^{\prime})$ for $L_{x}<30$ and $3*L_{x}/J(J^{\prime})$ for $L_{x}>34$.

\begin{figure}
\includegraphics[width=0.5\textwidth]{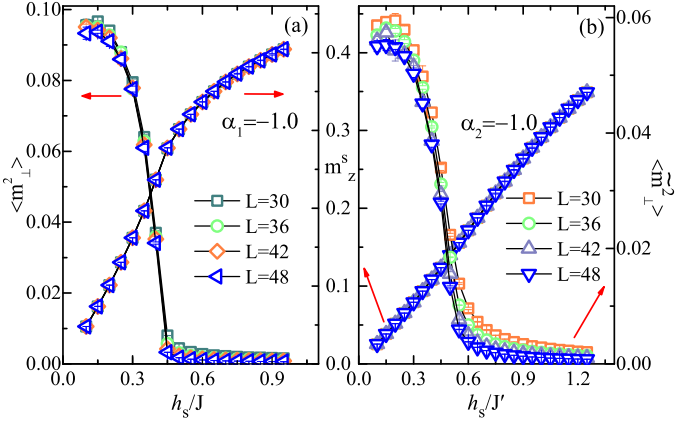}
\caption{(Color online) The transverse staggered magnetization square $\langle m_{\perp}^{2}\rangle$ ($\langle\tilde{m}_{\perp}^{2}\rangle$) and longitudinal staggered magnetization $m_{s}^{z}$ versus the staggered field $\emph{h}_{s}$ in the present system with coupling ratio (a) $\alpha_{1}=-1.0$ (Case A) and (b) $\alpha_{2}=-1.0$ (Case B). Here we take $L_{x}=L_{y}=L$. Most of the data are as accurate as $10^{-4}$.}
\label{staggered-stripe}
\end{figure}

Fig. \ref{staggered-stripe} presents the transverse staggered magnetization square $\langle m_{\perp}^{2}\rangle$ ($\langle\tilde{m}_{\perp}^{2}\rangle$) and longitudinal staggered magnetization $m_{s}^{z}$ as a function of staggered field $\emph{h}_{s}/J(J^{\prime})$ of the system for $\alpha_{1}=-1.0$ [Case A in Fig. \ref{staggered-stripe}(a)] and $\alpha_{2}=-1.0$ [Case B in Fig. \ref{staggered-stripe} (b)], where both are spin ordered in ground state in the absence of a magnetic field. It is observed that for both cases, with increasing the staggered magnetic field, $\langle m_{\perp}^{2}\rangle$ ($\langle\tilde{m}_{\perp}^{2}\rangle$) decreases from a finite value (around 0.1 for Case A and 0.06 for Case B) to sharply vanishing at a critical field $\emph{h}_{s}/J\simeq 0.45$ and $\emph{h}_{s}/J^{\prime}\simeq 0.50$ for Case A and Case B, respectively, where the in-plane QPT at critical fields appears to be of second-order, while $m_{s}^{z}$ increases almost linearly in the region of weak fields. This is understandable, as the staggered magnetic field is applied along the $z$ (out-of-plane) direction, with the increase of the field, the transverse magnetization in $xy$ plane will be gradually suppressed, while the longitudinal magnetization grows till saturation, as manifested in Fig.4. Recall that in the absence of an external field, the system with mixing F and AF bond couplings has an spin ordered ground state.

For the two cases, the behaviors of $\langle m_{\perp}^{2}\rangle$ and $\langle\tilde{m}_{\perp}^{2}\rangle$ look qualitatively similar, but the in-plane critical fields are somewhat different; $m_{s}^{z}$ behaviors slightly in a different way: Case A goes to magnetic saturation faster than Case B, because the former can be viewed as the antiferromagnetic zigzag spin chains coupled  ferromagnetically whereas the latter is formed by ferromagnetic zigzag spin chains coupled antiferromagnetically. In addition, the finite-size effect in Case B appears to be more obvious than Case A.

 \section{Scaling behavior in a staggered field}

 Binder ratios \cite{binder,kbinder,DPlandau} and spin stiffness \cite{sandvik} are proper quantities for investigating the critical features of the system. As $\emph{h}_{s}$ breaks the O(3) spin rotating symmetry, and the in-plane order parameter disappears at the critical point, we consider only the spin stiffness $\rho$ for simplicity. As mentioned in Sec.\uppercase\expandafter{\romannumeral3}, $\rho$ could be directly related to the superfluid density of a superconductor or superfluid \cite{jasnow,josephson,schultka,manousakis}, marking the occurrence of off-diagonal long-range order. According to the previous study \cite{Wang}, at the critical point it scales as $\rho\sim L^{2-d-z}$, where $d$ is the spatial dimension of the system, and $z$ is the dynamical exponent. Here we can set $z$ as 1, and measure $\rho L$ that is size-independent at the critical point.

\begin{figure}
\includegraphics[width=0.47\textwidth]{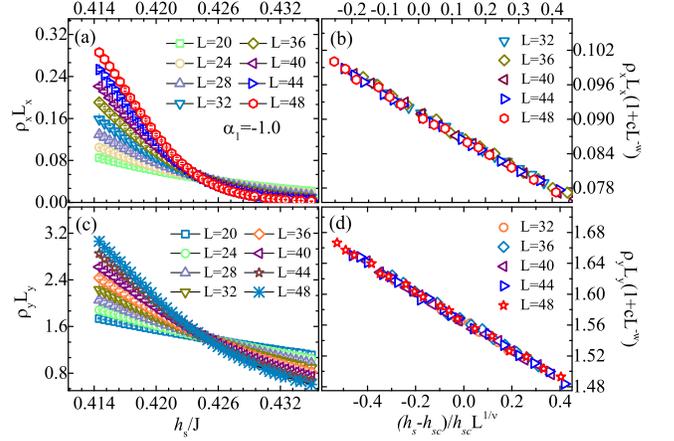}
\caption{(Color online) (a) $\rho_{x}L_{x}$ and (c) $\rho_{y}L_{y}$ as function of $\emph{h}_{s}/J$ of the system with coupling ratio $\alpha_{1}=-1.0$ near the critical point for various lattice sizes L=20, 24, 28, 32, 36, 40, 44 and 48 ; (b) and (d) are the corresponding data collapses for the finite-size scaling, where the data fall on a line, respectively, giving  a critical staggered field $\emph{h}_{sc}/J\simeq0.42373(4)$ and an exponent $\nu=0.677(2)$, which indicates that this is an O(3) universality transition. The errors for $\rho_{x}$ are mostly of $10^{-4}$ and for $\rho_{y}$ of $10^{-3}$.}
\label{fig-af-zigzag-data-collapse}
\end{figure}

For Case A with $\alpha_{1}=-1.0$, a staggered field $\emph{h}_{s}/J$ in [0.4145, 0.435] has been applied to the system on lattices $L_{x}=L_{y}=20$, 24, 28, 32, 36, 40, 44 and 48 with $\beta=100/J$ for $L_{x}<34$ and $3L_{x}/J$ for $L_{x}>34$. Figs. \ref{fig-af-zigzag-data-collapse} (a) and (c) present the field dependence of $\rho_{x}L_{x}$ and $\rho_{y}L_{y}$,  showing that the curves for different lattice sizes do intersect at about 0.424, which must be a critical point. To confirm this, we perform a finite-size scaling (FSS) by making data collapse analysis, as shown in Figs. \ref{fig-af-zigzag-data-collapse} (b) and (d), where all curves fall on a single almost straight line (see below for details).

\begin{figure}
\includegraphics[width=0.48\textwidth]{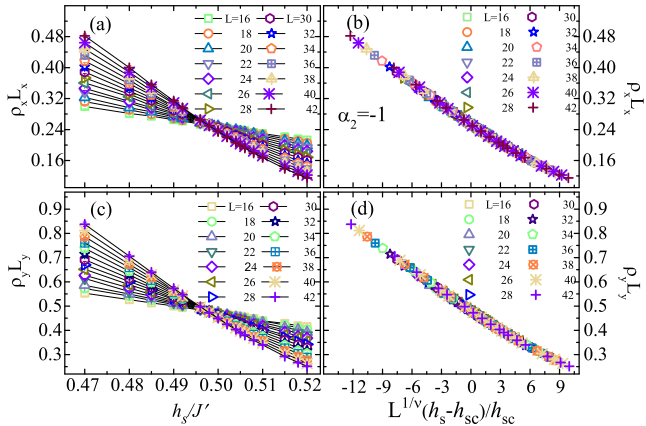}
\caption{(Color online) (a) $\rho_{x}L_{x}$ and (c) $\rho_{y}L_{y}$ as function of $\emph{h}_{s}/J^{\prime}$ of the system considered in Ref. \onlinecite{yizhen} with $\alpha_{2}=-1.0$ for various lattice sizes from L=16 to 42, where a crossing point is seen at the staggered field $\emph{h}_{s}/J^{\prime} \simeq 0.495$; (b) and (d) show the corresponding data collapse for the FSS fit, which gives $\emph{h}_{sc}/J^{\prime}\simeq0.497$ and $\nu\simeq0.693$, indicating that this QPT also belongs to the classical Heisenberg O(3) universality. The errors are at least $10^{-3}$.}
\label{fig-af-armchair-data-collapse}
\end{figure}

For Case B with $\alpha_{2}=-1.0$, we calculate $\rho_{x}L_{x}$ and $\rho_{y}L_{y}$ as function of staggered magnetic field on lattice sizes from L=12 to L=42 with $\beta=100/J^{\prime}$ for $L_{x}<34$ and $3L_{x}/J^{\prime}$ for $L_{x}>34$, as shown in Figs. \ref{fig-af-armchair-data-collapse}  (a) and (c), from which one may see that there is a crossing point at about $\emph{h}_{s}/J^{\prime}\simeq 0.495$, demonstrating that it may be a quantum critical point. The corresponding data collapses confirm this observation that all curves for different lattice sizes go to a single line [Figs. \ref{fig-af-armchair-data-collapse}  (b) and (d)]. It is also consistent with the vanishing points for $\langle\tilde{m}^{2}_{\perp}\rangle$ shown in Fig. \ref{staggered-stripe}, confirming the second-order QPT triggered by a staggered field. Here we would like to point out that in Ref. \onlinecite{yizhen} we discussed the scaling behaviors of $\rho_{x}L_{x}$ and $\rho_{y}L_{y}$ against the coupling ratio (here $|\alpha_{2}|$), where the QPT occurs at a critical coupling ratio, and the transition induced by the staggered magnetic field is left for our present study.

In the framework of renormalization group, the finite-size scaling plays as an essential role in studying the critical behavior near the transition point in finite-size systems \cite{fisher,brezin,barber,zinn}. In the vicinity of a critical point, the correlation length $\xi$ is divergent and, as the lattice size obeys $L\leq\xi$, some quantities exhibit power-law divergent behaviors with $\xi$ and could be expressed by a scaling function of the form $Q(t,L)=L^{\kappa/\nu}g(tL^{1/\nu})$,
where $\kappa$ is the critical exponent of Q and $\nu$ of $\xi$, $t$ is the reduced phase transition tuning parameter and $g(x)$ is a smooth function which asymptotically behaves as $g(x)\sim x^{-\kappa}$ for $x\rightarrow\infty$. Here for $\rho L$, $\kappa$ is zero and $t=(h_{s}-h_{sc})/h_{sc}$.

In Case A, the intersection points for different pairs $[L,L^{\prime}]$, where $L^{\prime}>L$, shift as $L$ enlarges, and a general scaling function under such conditions with extra corrections to $tL^{1/\nu}$ and $Q(t,L)$ \cite{beach} can be taken
\begin{eqnarray}
\begin{aligned}
(1+cL^{-\omega})Q(t,L)=g(tL^{1/\nu}+dL^{-\phi}).
\end{aligned}
\label{scaling-function}
\end{eqnarray}
The following scaling form will be more convenient
%\begin{spacing}{0.7}
\begin{eqnarray}
\begin{aligned}
(1+cL^{-\omega})Q(t,L)=a_{0}+a_{1}tL^{1/\nu}+a_{2}(tL^{1/\nu})^2,
\end{aligned}
\label{scaling-function1}
\end{eqnarray}
%\end{spacing}
where $c$, $\omega$, $a_{0}$, $a_{1}$ and $a_{2}$ are constants to be determined. For Case A only the correction $(1+cL^{-\omega})$ \cite{Ugerber} for $\rho_{x(y)}L_{x(y)}$ are included in the scaling functions.

For Case B, the form $Q(t,L)=L^{\kappa/\nu}g(tL^{1/\nu})$ works very well, and the scaling function is supposed to be polynomial of order two:
%\begin{spacing}{0.8}
\begin{eqnarray}
\begin{aligned}
Q^{\prime}(t,L)=a^{\prime}_{0}+a^{\prime}_{1}tL^{1/\nu}+a^{\prime}_{2}(tL^{1/\nu})^2,
\end{aligned}
\label{scaling-function2}
\end{eqnarray}
%\end{spacing}
where $a'_{0}$, $a'_{1}$, $a'_{2}$ and $\nu$ are constants independent of $L$.

The data are analyzed following the lines in Ref. \onlinecite{Wang}. We take thousands of copies of bootstrap resamplings of the raw data as the fitting data, and prepare the same amount of sets of initial fitting parameters in above functions as the input in fitting procedures, which are based on the nonlinear Levenberg-Marquardt optimization algorithm (LMOA) \cite{murray}. For Case A, $h_{s}/J$ in the area of $[0.4230, 0.4245]$ for lattice sizes $L_{x}=L_{y}=32$, 36, 40, 44, 48 are taken in the optimization procedure; while for Case B, all the data in Figs. \ref{fig-af-armchair-data-collapse} (a) and (c) are considered. The corresponding collapsed curves are shown in Figs. \ref{fig-af-zigzag-data-collapse} (b) and (d), and Figs. \ref{fig-af-armchair-data-collapse} (b) and (d), respectively.

Table. \ref{table1} presents the two sets of critical staggered magnetic field $h_{sc}$ and the exponent $\nu$ of correlation length for both cases. The values are determined by the lowest $\chi^{2}/DOF$ [e.g., the weighted sum of squares residual per degree of freedom (DOF)] for each single LMOA fitting. For case A, it gives $\emph{h}_{sc}/J=0.42373(4)$ and $\nu=0.677(2)$; and for case B, $\emph{h}_{sc}/J^{\prime}=0.49757(4)$ and $\nu=0.693(0)$. Considering the calculational errors, both cases are close to $\nu=0.7112(5)$ \cite{Massimo}, showing that these QPTs belong to the classical Heisenberg O(3) universality.

\begin{table}
\renewcommand{\arraystretch}{1.5}
\addtolength{\tabcolsep}{+4pt}
\caption{ The critical staggered magnetic field $\emph{h}_{sc}$ and the exponent $\nu$ of correlation length determined from  $\rho_{x(y)}L_{x(y)}$ for the present system (Case A) and the system (Case B) considered in Ref. \cite{yizhen}. }
\begin{tabular}[t]{c|cc|cc}
\hline \hline
 & Case A & & Case B &\\
\hline
&$\rho_{x}L_{x}$&$\rho_{y}L_{y}$ & $\rho_{x}L_{x}$&$\rho_{y}L_{y}$\\
$\emph{h}_{sc}/J(J^{\prime})$& 0.4233(1) &0.42373(4) &0.49757(4) &0.49734(8)\\
$\nu$&0.686(3) &0.677(2) & 0.693(0)&0.692(2) \\
\hline \hline
\end{tabular}
\label{table1}
\end{table}

\section{Phase diagram in a staggered field}

As shown in Fig. \ref{staggered-stripe}, when $\emph{h}_{s}<\emph{h}_{sc}$ both systems exhibit a staggered magnetization in $z$ direction and keep the corresponding stripe order in  $xy$ plane, that is, $m^{s}_{z}>0$ and $\langle m^{2}_{\perp}\rangle(\langle\tilde{m}^{2}_{\perp}\rangle)>0$. We coin the so-defined phase for $\emph{h}_{s}<\emph{h}_{sc}$ as the canted phase $\uppercase\expandafter{\romannumeral1}$ for Case A and the canted phase $\uppercase\expandafter{\romannumeral2}$ for Case B. As $\emph{h}_{s}>\emph{h}_{sc}$, only the out-of-plane staggered magnetization remains in both cases, say, $m^{s}_{z}>0$, and $\langle m^{2}_{\perp}\rangle(\langle\tilde{m}^{2}_{\perp}\rangle)=0$. We call such a phase for $\emph{h}_{s}>\emph{h}_{sc}$ as the $N\acute{e}el$ phase in a staggered field.

\begin{figure}
\includegraphics[width=0.47\textwidth]{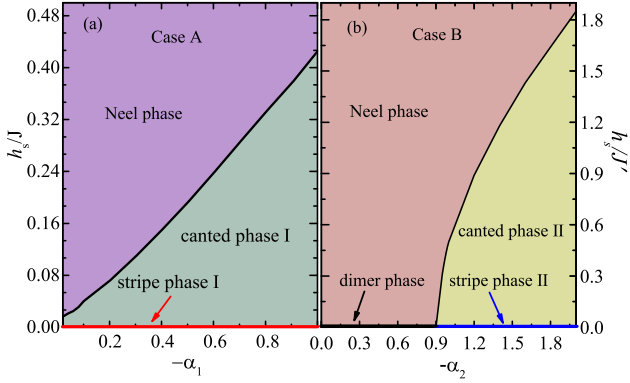}
\caption{(Color online) Phase diagram for the present system (Case A) and the system (Case B) considered in Ref. \onlinecite{yizhen} in the plane of the coupling ratio $\alpha_{1,2}$ versus staggered magnetic field $\emph{h}_{s}$.}
\label{phase-diagram}
\end{figure}

To draw a phase diagram, we inspect various coupling ratios for the two cases, and make use of the transition points in the curves of $m^{s}_{z}$, $\langle m^{2}_{\perp}\rangle$, $\langle\tilde{m}^{2}_{\perp}\rangle$, and $\rho L$ vs $h_{s}$ for each $\alpha_{1}$ or $\alpha_{2}$, forming the phase boundaries. In doing so, a schematic phase diagram in the plane of $h_{s}/J(J^{\prime})$ vs $-\alpha_{1(2)}$ is thus depicted in Fig. \ref{phase-diagram}.

For Case A, as shown in Fig. \ref{phase-diagram} (a), there are three phases, the stripe phase $\uppercase\expandafter{\romannumeral1}$ (where $m^{s}_{z}=0$ but $\langle m^{2}_{\perp}\rangle>0$), the canted phase $\uppercase\expandafter{\romannumeral1}$, and the $N\acute{e}el$ phase. The stripe phase $\uppercase\expandafter{\romannumeral1}$ always remains in the absence of a staggered field. When $h_{s}$ is increasing, the system immediately first enters into the canted phase $\uppercase\expandafter{\romannumeral1}$, and then enters into the $N\acute{e}el$ phase when $\emph{h}_{s}>\emph{h}_{sc}$.

For Case B, as shown in Fig. \ref{phase-diagram} (b), there are four phases, a dimerized phase, the stripe phase $\uppercase\expandafter{\romannumeral2}$, the canted phase $\uppercase\expandafter{\romannumeral2}$, and the $N\acute{e}el$ phase. Our previous study \cite{yizhen} shows that in the absence of a magnetic field, there is a phase transition at the critical point $\alpha_{2c}\simeq-0.93$ from a dimer phase to a stripe phase $\uppercase\expandafter{\romannumeral2}$ with a nonvanishing $\langle\tilde{m}^{2}\rangle=\frac{3}{2}\langle\tilde{m}^{2}_{\perp}\rangle$ but $m^{s}_{z}=0$. When the staggered field is increasing, the system in Case B immediately enters into the $N\acute{e}el$ phase for $\alpha<\alpha_{2c}$, while for $\alpha>\alpha_{2c}$, the system first enters into the canted phase  $\uppercase\expandafter{\romannumeral2}$ (where $m^{s}_{z}>0$ and $\langle\tilde{m}^{2}\rangle>0$) , and then enters into the $N\acute{e}el$ phase ($m^{s}_{z}>0$ but $\langle\tilde{m}^{2}\rangle=0$) when $\emph{h}_{s}>\emph{h}_{sc}$.

\section{Temperature dependence of susceptibility and specific heat in magnetic fields}

\begin{figure}
\includegraphics[width=0.5\textwidth]{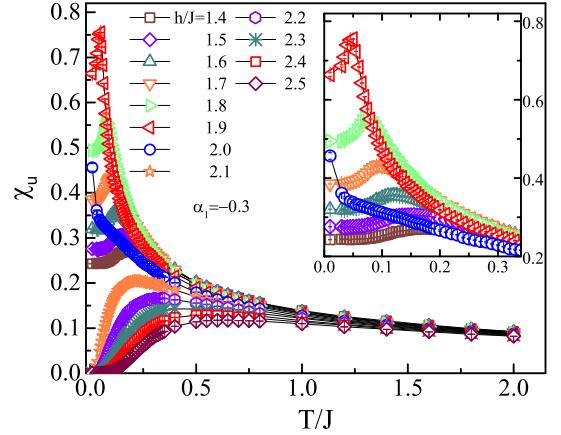}
\caption{(Color online) Temperature dependence of the susceptibility $\chi_{u}$ of the system with $\alpha_{1}=-0.3$ under various uniform magnetic fields. The inset shows $\chi_{u} (T)$ for $\emph{h}<\emph{h}_{c}$ and $T/J<0.34$. $\chi_{u}$ behaves differently in low temperature region: when $\emph{h}/J<2.0$, $\chi_{u}$ first goes to sharp peaks, and then decreases quickly; when $\emph{h}/J=2.0$, $\chi_{u}$ diverges as $T/J$ decreases, characterizing a critical point; and when $\emph{h}/J>2.0$, $\chi_{u}$ starts from a vanishing point at $T/J=0$ and forms a round peak. We find that as $\emph{h}_{c}-\emph{h}>0$ gets smaller, the peaks of $\chi_{u}$ get larger at small temperatures, and at a given temperature,  $\chi_{u}(\emph{h}<\emph{h}_{c})>\chi_{u}(\emph{h}>\emph{h}_{c})$. The accuracy here is as small as $10^{-4}$.}
\label{susceptibility-uniform-field}
\end{figure}

\begin{figure}
\includegraphics[width=0.5\textwidth]{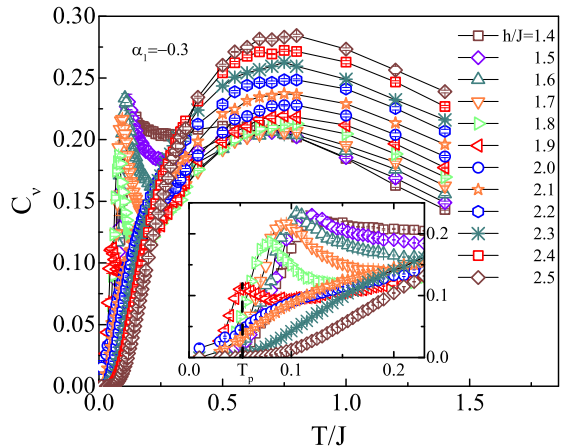}
\caption{(Color online) Temperature dependence of specific heat $C_{\nu}$ for the system with $\alpha_{1}=-0.3$ under different uniform fields. The inset is the low temperature part. Before the system is polarized, $C_{\nu}$ exhibits an extra peak at low temperature besides the round peak at relatively high temperature. Errors here are of $10^{-3}$.}
\label{specific-heat-uniform-field}
\end{figure}

\begin{figure}
\includegraphics[width=0.5\textwidth]{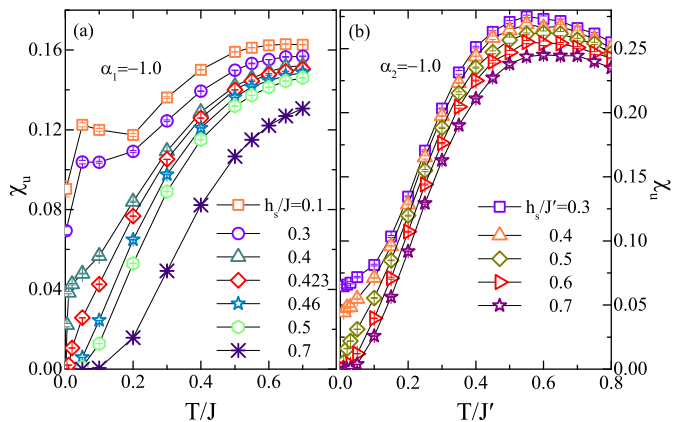}
\caption{(Color online) Temperature dependence of susceptibility $\chi_{u}$ of the system with (a) $\alpha_{1}$=-1.0 and (b) $\alpha_{2}$=-1.0 under various staggered magnetic fields. The order for the error bars are $10^{-3}$.}
\label{susceptibility-staggered-field}
\end{figure}

In this section, we study the temperature dependence of the susceptibility $\chi_{u}$ and specific heat $C_{\nu}$  under different magnetic fields.  The results are given in Figs. \ref{susceptibility-uniform-field}, \ref{specific-heat-uniform-field} and \ref{susceptibility-staggered-field}, respectively.

In subsection \uppercase\expandafter{\romannumeral4} A, it shows that for $\alpha_{1}=-0.3$ the system is polarized when $\emph{h}/J\geq2.0$, and when $\emph{h}/J<2.0$, the system stays in a canted stripe state with $m^{s}_{\bot}>0$ and $m_{z}>0$. In Fig. \ref{susceptibility-uniform-field},  one may see that for $\emph{h}/J<2.0$, $\chi_{u}$ increases from a finite value with increasing temperature, which becomes larger with the increase of  the magnetic field, and after undergoing a maximum it decreases quickly at low temperature; when $\emph{h}$ gets higher, the peak is sharper; as $\emph{h}$ is close to the critical point, $\chi_{u}$ decays almost exponentially. When $\emph{h}/J=2.0$, $\chi_{u}$ diverges as $T$ decreases. For $\emph{h}/J>2.0$, $\chi_{u} (T)$ is suppressed by the magnetic field, leading to all curves are below those of $\emph{h}/J<=2.0$, showing the system enters into a different state. At high temperature, all curves coincide with each other owing to the domination of thermal fluctuations. At low temperature, when the system is partially polarized, $\chi_{u}$ is influenced mainly by the transverse quantum fluctuations in $xy$ plane, and closer to the critical point, stronger the quantum fluctuations and higher $\chi_{u}$.

Fig. \ref{specific-heat-uniform-field} shows temperature dependence of specific heat $C_{\nu}$ of the system with $\alpha_{1}=-0.3$ under uniform fields. It can be observed that besides a round peak as that for $\emph{h}/J>2.0$, in the range of $1.4<\emph{h}/J<2.0$, $C_{\nu}$ exhibits a sharp peak at lower temperature $T_{p}$, as shown in the inset of Fig. \ref{specific-heat-uniform-field}. When $\emph{h}$ increases, $T_{p}$ and $C_{\nu}(T_{p})$ decrease. This could be understood in the following way. As indicated in Fig. \ref{interaction-pattern}, the spins along the zigzag chain form a continuous antiferromagnetic arrangement, dividing the system into two sublattices and producing two sets of degenerate spin wave spectra: $\hbar\omega_{k}=f(J,J^{\prime},S,\gamma_{k})$, where $S$ is the spin on each site, and $\gamma_{k}$ is the static structure factor. A uniform field $\emph{h}$ would split this overlapping spectra with a shift, $\hbar\omega_{k}^{\pm}=f(J,J^{\prime},S,\gamma_{k})\pm\emph{h}$, resulting in that two modes of low-lying excitations cause two minimums \cite{Klumper,Su} in $C_{\nu}$. The part of $C_{\nu}$ contributed by lower frequency mode increases faster. $\emph{h}$ enlarges the difference of the increasing tendency between $\hbar\omega_{k}^{+}$ and $\hbar\omega_{k}^{-}$, making $C_{\nu}$ steeper and $T_{p}$ smaller for larger $\emph{h}<\emph{h}_{c}$ at low temperature. When $\emph{h}>\emph{h}_{c}$, the system enters into the polarized state, and $C_{\nu}$ versus $T/J$ exhibit only one round peak at a relatively higher temperature.

In Fig. \ref{susceptibility-staggered-field}, it depicts the temperature dependence of susceptibility $\chi_{u}$ for systems with $\alpha_{1}=-1.0$  and $\alpha_{2}=-1.0$ under different staggered magnetic fields, which differ from those in a uniform magnetic field. We showed that a staggered field can induce a quantum phase transition in Sec. \uppercase\expandafter{\romannumeral5}, which eliminates the off-diagonal long range order in the $xy$ plane at $\emph{h}_{sc}/J\simeq0.423$ for $\alpha_{1}=-1.0$ and $\emph{h}_{sc}/J^{\prime}\simeq0.5$ for $\alpha_{2}=-1.0$. Figs. \ref{susceptibility-staggered-field} (a) and (b) illustrate that, as $\emph{h}_{s}<\emph{h}_{sc}$, $\chi_{u}$ would start from a non-vanishing value, while as $\emph{h}_{s}\geq\emph{h}_{sc}$, the susceptibility $\chi_{u}$ goes to zero at $T\rightarrow 0$, revealing that the system in this situation enters into distinct phases under different staggered magnetic fields, consistent with the observation in Fig. \ref{phase-diagram}.

\section{Summary}

The spin-1/2 Heisenberg model with AF and F mixing interactions on honeycomb lattice has been studied by means of the continuous imaginary-time QMC with the worm update algorithm in uniform and staggered magnetic fields. It is found that so long as the F coupling on armchair bonds is tuned on, the system (Case A) is immediately crossover smoothly from 1D disordered AF zigzag spin chains to a stripe ordered 2D phase with $\langle m^{2}_{s}\rangle>0$. This is in contrast to the system considered in Ref. \onlinecite{yizhen} where the F interactions are presumed on zigzag bonds and AF interactions on armchair bonds. In this latter system (Case B), upon tuning on the F interactions on zigzag bonds, the system is crossover smoothly from a disordered dimerized phase to a stripe ordered 2D phase.

In the presence of uniform or staggered magnetic fields, it is shown that for a given coupling ratio (e.g. $\alpha_{1}=-0.3$ in a uniform field, and $\alpha_{1,2}=-1.0$ in a staggered field), with increasing the external magnetic fields, the system enters smoothly into a spin canted phase from a stripe order phase, and then undergoes a QPT into an out-of-plane polarized phase or $N\acute{e}el$ phase. This is also true for the system in Case B for the coupling ratio beyond a critical value (satisfying $-\alpha_{2}>0.93$). The whole phase diagrams in the plane of coupling ratio and staggered magnetic field for the systems in Case A and Case B are obtained. In Case A, there are three phases, including stripe order phase, canted phase and $N\acute{e}el$ phase, while in Case B, there are four phases, say, dimerized phase, stripe phase, canted phase and $N\acute{e}el$ phase.

In addition, by exploring the spin stiffness, the scaling behaviors in a staggered field for both systems are also discussed. The finite-size scaling analysis gives that the exponent $\nu$ of correlation length is $\nu$=0.677(2) and 0.693(0) for Case A and Case B, respectively, which is very close to  $\nu=0.7112$ of classical Heisenberg O(3) universality, indicating that both systems fall into the O(3) universality. Besides, the scaling functions are different from the two systems.

The temperature dependence of susceptibility $\chi_{u}$ and specific heat $C_{\nu}$ have been studied for the system with $\alpha_{1}=-0.3$ under various uniform fields. When the system stays in the canted stripe phase as $\emph{h}/J<2.0$, at low temperature the partially polarized spins have a nonzero value of $\chi_{u}$ in the ground state and a sharp peak of $\chi_{u}(T)$ appears at low temperature, and meanwhile, the specific heat $C_{\nu}$ also presents a sharp peak starting from a vanishing value when $T/J\rightarrow0$. As $\emph{h}/J\geq2.0$, the polarized ferromagnetic state does not display such features because of nondegenerate spin wave spectra which could be separated by the uniform magnetic field. The behaviors of $\chi_{u}$ versus $T/J(J^{\prime})$ for $\alpha_{1}=-1.0$ and $\alpha_{2}=-1.0$ under different staggered fields are consistent with the phase diagram presented in Sec. \uppercase\expandafter{\romannumeral5}.

The present study shows that the competition among mixing interactions, external probes such as staggered magnetic field and temperature as well as the topology of the lattice, altogether, results in more complex phenomena in quantum many-body systems. Our results would also be helpful for further understanding the physical properties and scaling behaviors in 2D magnetic materials with mixing AF and F interactions.

\acknowledgments

The authors acknowledge W. Li, B. Xin, X. Yan, and Z. C. Wang for helpful discussions. This work was supported in part by the MOST of China (Grant  No. 2013CB933401), the NSFC (Grant No. 11474279), and the Strategic Priority Research Program of the Chinese Academy of Sciences (Grant No. XDB07010100).

\end{document}